\documentclass[aps,prd,twocolumn,eqsecnum,showpacs,preprintnumbers,superscriptaddress]{revtex4}
\usepackage{graphicx}

\newcommand{\be}{\begin{equation}}
\newcommand{\ee}{\end{equation}}
\newcommand{\ba}{\begin{eqnarray}}
\newcommand{\ea}{\end{eqnarray}}
\newcommand{\ep}{\varepsilon}
\newcommand{\nn}{\nonumber}
\newcommand{\ra}{\rightarrow}
\newcommand{\lra}{\leftrightarrow}

\newcommand{\ita}{\textit}

\newcommand{\nc}{N_C} 
\newcommand{\nl}{n_{l}}
\newcommand{\zt}{z_t}
\newcommand{\zu}{z_u}
\newcommand{\hspn}{{\hspace{-0.6mm}}}
\newcommand{\cf}{{C^{}_F}}

\begin{document}

\preprint{MZ--TH/08--05}
\preprint{DESY 08--008}

\title{Next-to-next-to-leading order 
${\cal O}(\alpha_s^4)$ results for heavy
quark pair production in quark--antiquark collisions:
The one-loop squared contributions}


\author{J.\ G.\ K\"{o}rner}
\email[Electronic address:]{koerner@thep.physik.uni-mainz.de}
\affiliation{Institut f\"{u}r Physik, Johannes
Gutenberg-Universit\"{a}t, D-55099 Mainz, Germany}

\author{Z.\ Merebashvili}
\email[Electronic address:]{zakaria.merebashvili@desy.de}
\affiliation{II. Institut f\"{u}r Theoretische Physik,
Universit\"{a}t Hamburg, Luruper Chaussee 149, 22761 Hamburg, Germany}

\author{M.\ Rogal}
\email[Electronic address:]{Mikhail.Rogal@desy.de}
\affiliation{Deutsches Elektronen-Synchrotron DESY, Platanenallee 6, D-15738
Zeuthen, Germany}

\date{\today}

\begin{abstract}
We calculate the next-to-next-to-leading order 
${\cal O}(\alpha_s^4)$ one-loop squared corrections
to the production
of heavy quark pairs in quark-antiquark annihilations.
These are part of the
next-to-next-to-leading order 
${\cal O}(\alpha_s^4)$ radiative QCD
corrections to this process.
Our results, with the full mass dependence retained,
are presented in a closed and very compact form, in the dimensional 
regularization scheme. We have found very intriguing factorization properties 
for the finite part of the amplitudes.
\end{abstract}


\maketitle

\section{\label{intro}Introduction}

There was recently much activity in the phenomenology of hadronic heavy 
quark pair production in connection with the Tevatron and the CERN 
Large Hadron 
Collider (LHC) which will have its startup this year. There will be 
much experimental effort dedicated to the discovery of the Higgs boson.
There will also be studies of the copious production of top quarks and other
heavy particles, which also serve as a background to Higgs
boson searches as well as to possible new physics beyond the standard
model. Therefore, it is mandatory to reduce the 
theoretical uncertainty in phenomenological 
calculations of heavy quark production processes as much as possible. 

Several years ago the next-to-next-to-leading order (NNLO) contributions to 
hadron production were 
calculated  by several groups in massless QCD (see e.g. \cite{Glover} and 
references therein).
The
completion of a similar program for processes that involve massive
quarks requires much more dedication and work since the inclusion of
an additional mass scale dramatically complicates the whole
calculation. 

Until very recently there was the belief that the next-to-leading order (NLO) 
description of
heavy charm and bottom production in hadronic collisions considerably 
underestimates the experimental results. In recent, more refined analyses
\cite{Cacciari:2003uh,Bernd1,Bernd2} it was shown that a NLO analysis does 
in fact properly describe the latest charm and bottom quark 
production data \cite{CDF}. The authors of \cite{Cacciari:2003uh} and 
\cite{Bernd1,Bernd2} deal very differently with the 
problem of large mass logarithms which constitute the central problem 
in the heavy quark phenomenology. Data on top quark 
pair production also agrees with the NLO prediction within theoretical and
experimental errors 
(see e.g \cite{Chakraborty:2003iw}). In all of these NLO calculations there 
remains, among others, the problem 
that the renormalization and factorization scale dependence of the NLO
calculations render the theoretical results quite uncertain. This calls
for a NNLO calculation of heavy quark production in hadronic collisions
which is expected to considerably reduce the scale dependence of the
theoretical prediction. 

At the lower energies of Tevatron II, top quark pair production is dominated
by $q {\bar q}$ annihilation (85 \%). The remaining 15\% come from
gluon fusion. At the higher energy LHC, gluon fusion dominates
the production process (90 \%) with 10 \% left for $q {\bar q}$ annihilation
(percentage figures from \cite{Chakraborty:2003iw}). This shows that both
$q {\bar q}$ annihilation and gluon fusion have to be accounted for in the 
calculation of top quark pair production.

In general, there are
four classes of contributions that need to be calculated for the
NNLO corrections to the hadronic production of heavy quark pairs.
The first class involves the pure two-loop contribution, which has to be
folded with the leading order (LO) Born term. The second class of diagrams
consists of the so-called one-loop squared contributions (also called 
loop--by--loop contributions)
arising from the product of one-loop virtual matrix elements. This is the
topic of the present paper. Further, there are the
one-loop gluon emission contributions that are folded with the one--gluon
emission graphs. Finally, there are the squared two-gluon emission 
contributions that are purely of tree--type.

Bits and pieces of the NNLO calculation are now being assembled. The recent
two--loop calculation of the heavy quark vertex form factor 
\cite{bernreuther05a} can be used as one of the many building blocks in the 
first class of processes. In this context we would also like to mention the 
recent work \cite{Moch} on the NNLO calculation of
two-loop virtual amplitudes performed in the domain of high energy 
asymptotics, where the heavy quark mass is small compared to the other large
scales. In this calculation mass power
corrections are left out, and only large mass logarithms and finite terms
associated with them are retained. 
The authors of the 
present paper have been involved in a systematic effort
to calculate all the contributions from the second class of processes, the
one--loop squared contributions. We
shall describe the present status of this program in the next paragraph.
In the work \cite{Stefan} the full, exact NLO corrections to $t\bar{t}+$jet 
are presented. When integrating over the full phase space of the jet 
(or gluon), this 
calculation can be turned into a NNLO calculation of heavy hadron production
of the third class. To our knowledge there does not exist a complete 
calculation of the fourth class of processes, the squared two-gluon 
emission contributions. 

Let us briefly describe the status of our effort to calculate the one--loop
squared contributions for the second class of processes.
The highest singularity in the one--loop amplitudes from infrared (IR) and
mass singularities (M) is, in general, proportional to $(1/\ep^2)$. This in turn
implies that the Laurent series expansion of the one--loop amplitudes
has to be taken up to ${\cal O}(\ep^2)$ when calculating the one-loop squared
contributions. In fact, it is the ${\cal O}(\ep^2)$ terms in the Laurent
series expansion that really complicate things \cite{KMR} since the
${\cal O}(\ep^2)$ contributions in the one-loop amplitudes involve a
multitude of multiple polylogarithms of maximal weight and depth 4 
\cite{KMR1}. All scalar master integrals needed in this calculation have
been assembled in \cite{KMR,KMR1}. Reference \cite{KMR} gives the results in terms
of so-called $L$ functions, which can be written as one-dimensional integral 
representations involving products of log
and dilog functions, while \cite{KMR1} gives the results in terms of multiple 
polylogarithms. The divergent and finite terms of the one--loop amplitude
$q{\bar q} \to Q{\bar Q}$ were given in \cite{KM}. The remaining 
${\mathcal O}(\ep)$ and ${\mathcal O}(\ep^{2})$ amplitudes have
been written down in \cite{KMR2}. Squaring the one--loop amplitudes leads to 
the results of the present paper. In a recent work \cite{gamgam}
we have presented closed-form, one-loop squared results for heavy quark 
production in the fusion of real photons.

\begin{figure*}
\includegraphics{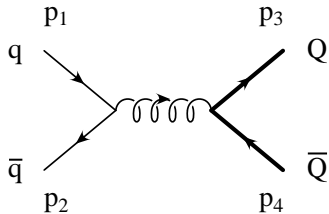}
\caption{\label{fig:born}
The lowest order Feynman diagram representing light quark--antiquark annihilation.
Normal solid lines represent the light quarks,
the curly line represents the gluons and the
thick solid lines correspond to the heavy quarks.}
\end{figure*}


In this paper we report on a calculation of the NNLO one--loop
squared matrix elements for the process
$q\bar{q} \to Q \overline{Q}$. The
calculation is carried out in the dimensional
regularization scheme \cite{DREG} with space-time dimension $n=4-2\ep$.
In sequels to this paper we shall present results on the square of 
hadroproduction amplitudes originating from the gluon fusion subprocess 
$gg \to Q \overline{Q}$ and
photoproduction amplitudes $\gamma g \to Q \overline{Q}$.

In our presentation we shall
make use of our notation for the coefficient functions of the relevant
scalar one-loop master integrals calculated up to ${\cal O}(\ep^2)$ in 
\cite{KMR}.
For the case of gluon-gluon and quark-antiquark collisions, one needs
all the scalar integrals derived in \cite{KMR}, e.g.
the one scalar one-point
function $A$; the five scalar two--point functions $B_1$, $B_2$, $B_3$,
$B_4$, and $B_5$; the six scalar
three--point functions $C_1, C_2, C_3, C_4, C_5$, and $C_6$; and the three
scalar four-point 
functions $D_1, D_2$, and $D_3$.
Taking the {\it complex} scalar four-point function $D_2$ as an example, we 
define successive coefficient functions $D_2^{(j)}$ for the Laurent
series expansion of $D_2$.
One has
\begin{eqnarray}
\label{Dexp}
D_2&\!\!=\!\!&i C_\ep(m^2)\Big\{\frac{1}{\ep^{2}}D_2^{(-2)}+\frac{1}{\ep}D_2^{(-1)} + 
D_2^{(0)} + \ep D_2^{(1)} \nonumber \\ 
&& + \ep^2 D_2^{(2)} + {\mathcal O}(\ep^3) \Big\},
\end{eqnarray}
where $C_{\ep}(m^2)$ is defined by
\be
\label{ceps}
C_{\ep}(m^2)\equiv\frac{\Gamma(1+\ep)}{(4\pi)^2}
\left(\frac{4\pi\mu^2}{m^2}\right)^\ep .
\ee
We use this notation for both the real and the imaginary parts of $D_2$,
i.e. for ${\rm Re}D_2$ and ${\rm Im}D_2$.
Similar expansions hold for the scalar one--point
function $A$, the scalar
two--point functions $B_i$, the scalar three--point functions $C_i$, 
and the remaining four-point functions $D_{i}$.
The coefficient functions of the various Laurent
series expansions were given in \cite{KMR} in the form of so--called
$L$ functions, and in \cite{KMR1} in terms of multiple 
polylogarithms of maximal weight and depth 4. It is then a matter of 
choice which of the two representations are used for the numerical
evaluation. The numerical evaluation of the $L$ functions in terms of their 
one--dimensional integral representations is quite straightforward using 
conventional integration routines, while there exists a very efficient 
algorithm to numerically evaluate multiple polylogarithms 
\cite{Vollinga:2004sn}.  

Let us summarize the main features of the scalar master integrals. The master
integrals $A,B_{1},B_{3},B_{4},C_{2},C_{3}$, and $D_{3}$ are purely real, whereas
$B_{2},B_{5},C_{1},C_{4},C_{5},C_{6},D_{1}$, and $D_{2}$ are truly complex. From 
the form $(AB^{\ast}+BA^{\ast})=2({\rm Re}A\,{\rm Re}B+{\rm Im}A\,{\rm Im}B)$ 
it is clear that the imaginary parts of the master integrals
must be taken into account in the one-loop squared contribution. The master 
integrals 
$B_{2},B_{5},C_{1},C_{4},C_{5}$, and $C_{6}$ are $(t\lra u)$ symmetric, 
where the kinematic variables $t$ and $u$ are defined in Sec.~\ref{notation}.
 
The paper is organized as follows. Section~\ref{notation} contains an
outline of our general approach and discusses renormalization procedures.
Section~\ref{nlo} presents NLO results 
for the quark-antiquark annihilation subprocess.
In Sec.~\ref{singular} one finds a discussion of the singularity
structure of the NNLO squared matrix element for the quark-antiquark
annihilation subprocess.
In Sec.~\ref{finite} we discuss the structure of the finite part of our
result.
Our results are summarized in Sec.~\ref{summary}.
In the Appendix we write down expressions for the building blocks of that
part of the finite result that originates from the square of box diagrams.


\section{\label{notation}
NOTATION
}


The heavy flavor hadroproduction proceeds through two partonic
subprocesses: gluon fusion and light quark-antiquark annihilation. The
first subprocess is the most challenging one in QCD from a technical point
of view. It has three production topologies already at the Born level. Here
we consider the second subprocess where there is only one topology 
at the Born term level (see Fig.~\ref{fig:born}). Irrespective of the partons 
involved, the general kinematics is, of course, the same in both processes. In 
particular, for the quark-antiquark annihilation, Fig.~\ref{fig:born}, we have
\be
\label{qq}
q(p_1) + \bar{q}(p_2) \ra Q(p_3) + \overline Q(p_4),
\ee


The momenta directions correspond to the physical configuration; e.g.
$p_1$ and
$p_2$ are ingoing whereas $p_3$ and $p_4$ are outgoing.
With $m$ being the heavy quark mass, we define
\ba
\nn
&s\equiv (p_1+p_2)^2, \qquad  t\equiv T-m^2 \equiv
(p_1-p_3)^2-m^2,&
\\
&u\equiv U-m^2\equiv (p_2-p_3)^2-m^2, &
\ea
so that the energy-momentum conservation reads $s+t+u=0$.

We also introduce the overall factor
\be
\label{common}
\mathcal C = \left( g_s^4 C_{\ep}(m^2) \right)^2,
\ee
where $g_s$ is the renormalized strong coupling constant 
and $C_\ep(m^2)$ is defined in (\ref{ceps}).


As shown e.g. in \cite{KM,KMR2} the self--energy
and vertex diagrams contain ultraviolet (UV) and infrared and collinear
(IR/M) poles even after
heavy mass renormalization. The UV poles need to be regularized.

Our renormalization procedure is carried out as follows: when dealing
with massless
quarks we work in the $\overline{\rm MS}$ scheme, while heavy quarks are
renormalized in the on--shell scheme, where the heavy quark mass is the pole
mass.
For completeness we list the set of
one-loop renormalization constants that we have used:
\ba
\nn   &&
Z_1 = 1 + \frac{g^2_s}{\ep} \frac{2}{3} \left\{ (N_C - n_{l}) C_{\ep}(\mu^2)
                   - C_{\ep}(m^2) \right\},     \\
\nn   &&
Z_m = 1 - g^2_s C_F C_{\ep}(m^2) \frac{3-2\ep}{\ep (1-2\ep)},    \\
&&
Z_2 = Z_m,   \\
\nn   &&
Z_{1F} = Z_2 - \frac{g^2_s}{\ep} N_C C_{\ep}(\mu^2),    \\
\nn   &&
Z_{1f} = 1 - \frac{g^2_s}{\ep} N_C C_{\ep}(\mu^2),    \\
\nn   &&
Z_3 = 1 + \frac{g^2_s}{\ep} \left\{ (\frac{5}{3} N_C - \frac{2}{3}
n_{l}) C_{\ep}(\mu^2) - \frac{2}{3} C_{\ep}(m^2)\right\}
\\  \nn &&  \qquad
= 1 + \frac{g^2_s}{\ep}\left\{ (\beta_0 - 2N_C) C_{\ep}(\mu^2) -
\frac{2}{3} C_{\ep}(m^2)\right\},      \\
\nn   &&
Z_g =  1 - \frac{g^2_s}{\ep}\left\{ \frac{\beta_0}{2} C_{\ep}(\mu^2) -
\frac{1}{3} C_{\ep}(m^2)\right\}.
\ea
with $\beta_0=(11 N_C - 2 n_{l})/3$. $n_{l}$ is the number of light quarks,
$C_F=4/3$, and $N_C=3$ is the number of colors. The arbitrary mass 
scale $\mu$ is the scale at which the renormalization is carried out. The above
renormalization constants are as follows: $Z_1$ for the three-gluon vertex, 
$Z_m$ for the heavy quark mass, $Z_2$ for the heavy quark wave function, 
$Z_{1F}$ for the $(Q\overline Qg)$ vertex, 
$Z_{1f}$ for the $(q\overline qg)$ vertex,
$Z_3$ for the gluon wave function, and $Z_g$ for the strong coupling 
constant $\alpha_{s}$. Note that $Z_1$ is not actually needed
in the present application, but we have presented it for completeness.
For the massless quarks there is no mass and wave function renormalization.

\begin{figure*}
\includegraphics{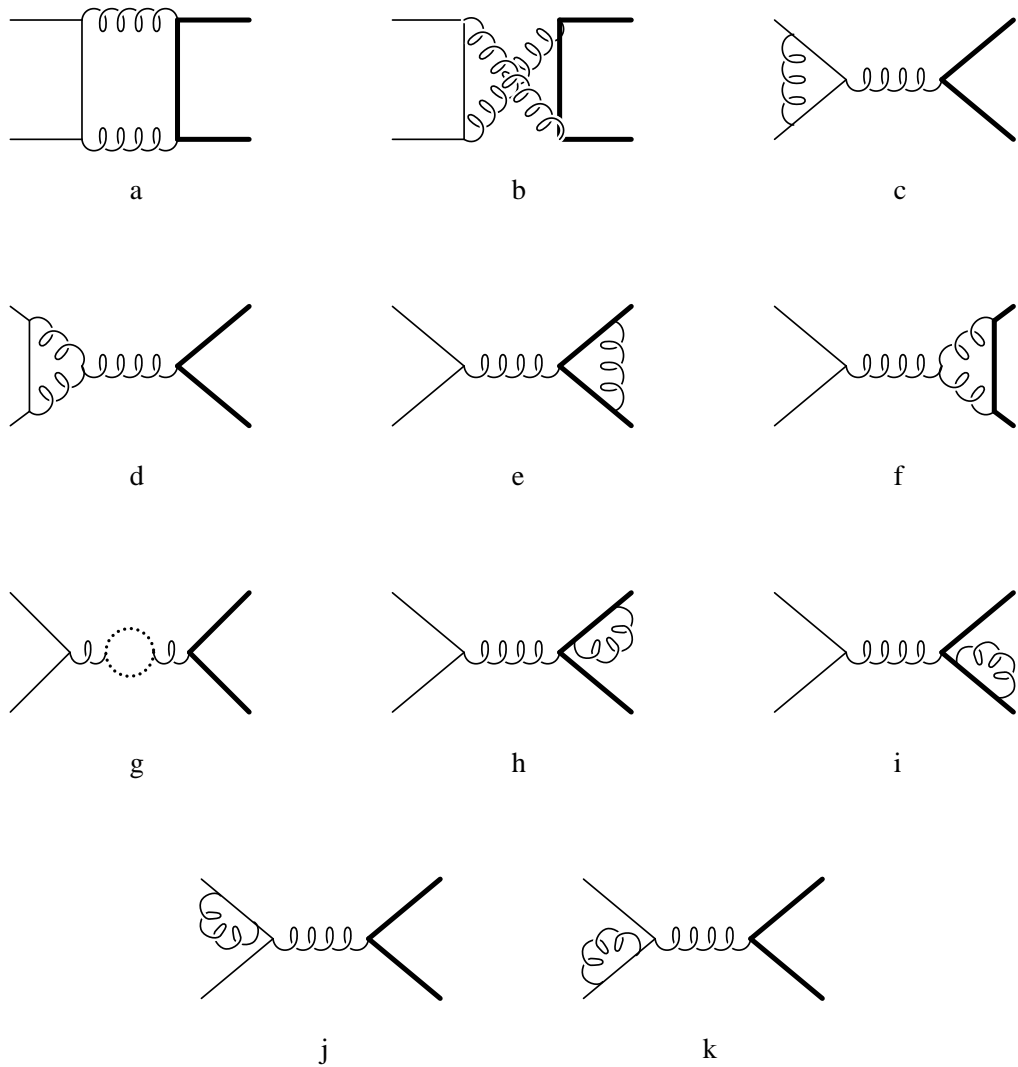}
\caption{\label{fig:nlo}
One-loop Feynman diagrams contributing to the subprocess $q\bar{q}\to Q\overline{Q}$.
The loop with the dotted line in (g) represents the gluon, ghost, 
and light and heavy quarks.}
\end{figure*}

The above coefficients (except for $Z_g$) are needed if one 
renormalizes graph by graph.
However, one could choose another route.
From the field-theoretical point of view, the renormalized matrix
element is obtained from the unrenormalized one by
\be
M_{\rm ren} = \prod_n Z_{f_n}^{-1/2} M_{\rm bare} (g_{\rm bare} \ra Z_g
g_s, m_{\rm bare} \ra Z_m m_r),
\ee
where $Z_{f_n}$ are the wave function renormalization constants for
all the external on-shell particles under consideration. 
If one formally expands $M_{\rm bare}$ (e.g. $M_{\rm bare}=M_0+g^2_{s}
M_1+\ldots)$ and the
renormalization parameters $Z_{f_n}$
as a series of powers in the coupling constant to the requisite order, one
arrives at the one-loop order result 
\ba
\nn   &
M_{\rm 1,ren} = \prod_n Z_{f_n,1}^{-1/2} M_{0} (g_{\rm bare} \ra
Z_g g_s, m_{\rm bare} \ra Z_m m_r)     & \\
& + g^2_{s} M_1(g_s,m_r),  &
\ea
where now the $Z_{f_n,1}$ correspond to the one--loop renormalization
constants for the external particles.
In our case one has
$Z_{f_1,1}=Z_{f_2,1}=1$ and $Z_{f_3,1}=Z_{f_4,1}=Z_{2}$. 
Thus, one could apply inverse wave function
renormalization for external legs and then replace the
bare coupling constant $g_{\rm bare} \ra Z_g g_s$ (as the mass 
parameter $m$ does not explicitly enter the leading order Born 
term matrix element, it is not renormalized at that order).
We have verified that, in both ways, we arrive at the same
renormalized result.

In order to fix our normalization we write down the differential cross section
for $q {\bar q} \to Q \overline{Q}$ in
terms of the squared amplitudes $|M|^2$.
One has
\be
d\sigma_{q \bar{q} \rightarrow  Q \overline{Q} }=
\frac{d({\rm PS})_2}{2 s}  \frac{1}{4 N_C^2}
|M|^2_{q \bar{q} \rightarrow  Q \overline{Q} } \, ,
\ee
where the $n$--dimensional two--body phase space is given by
\be
d ({\rm PS})_2=
\frac{m^{-2\ep}}{8 \pi s} \frac{(4\pi)^{\ep}}{\Gamma (1-\ep)}
\left( \frac{tu-sm^2}{sm^2} \right)^{-\ep} \delta (s+t+u) dtdu
\ee
and we explicitly show flux $(4p_1p_2)^{-1} = (2s)^{-1}$, 
initial quark and antiquark spin $(2s_{f}+1)^{-2}=1/4$, and color $N_{C}^{-2}$
averaging factors. Then, at the leading Born term order for 
$q \bar{q} \rightarrow  Q \overline{Q}$,
we have
\be
\label{lo}
\frac{1}{g_s^4} |M|^2_{\rm LO} =
16 \left( \frac{t^2 + u^2}{s^2} + 2 \frac{m^2}{s} - \ep\right)   \equiv B .
\ee


\section{\label{nlo}
Next-to-leading order result
}


Folding the one--loop matrix elements depicted in Fig.~\ref{fig:nlo} with the 
LO Born term, Fig.~\ref{fig:born}, one obtains the virtual part of the NLO result.
Although NLO virtual corrections to heavy flavor hadroproduction were calculated 
before for the $q{\bar q}\to Q{\bar Q} $ case, one cannot find explicit 
results for this subprocess in the literature. We have therefore recalculated 
the virtual NLO contribution to $q{\bar q}$--annihilation. In fact, we have
calculated the virtual NLO results up to ${\mathcal O}(\ep^2)$. As it turns out,
the expressions for the NLO virtual $\ep^1$ and $\ep^2$ contributions
considerably simplify the presentation of the corresponding NNLO results, in as
much as they appear as important building blocks in the NNLO results.

\begin{widetext}
First, we write down a few abbreviations that we shall use throughout the 
paper:
\ba   &
\beta=\sqrt{1-4m^2/s}, \qquad   D=m^2 s - t u, & \\
\nn   &
z_2=s + 2 t, \qquad     z_{2u}=s + 2 u,   \qquad
z_t=2 m^2 + t,       \qquad    z_u=2 m^2 + u.  &
\ea

Note that $D$ is {\it not} the space--time dimension. 
We further define the functions:
\ba  &&
F_1^{(j)} = \frac{2}{9} (n_{l} + 1) + \frac{28 N_C}{9} - \frac{N_C}{\beta^2}
   - B_2^{(j)} \left( 3 C_F - \frac{3}{2} N_C + 1 - \frac{\beta^2}{3} \right) \\
\nn &&   \qquad \qquad 
   - B_5^{(j)} \left( 3 C_F - \frac{5 N_C}{3} + \frac{2 n_{l}}{3}  
           - \frac{N_C}{2\beta^2} \right) 
   + C_1^{(j)} N_C \frac{m^2}{\beta^2} 
   - \left\{C_4^{(j)} s - C_6^{(j)} (2 m^2 - s) \right\} (2 C_F - N_C) ,
\ea
\ba  &&
F_2^{(j)} = 2 \left( s \beta^2 (2 C_F - N_C) - 12 m^2 N_C \right) 
     - B_2^{(j)} s \beta^2 (2 C_F - N_C) + B_5^{(j)} (8 m^2 + s) N_C 
     + C_1^{(j)} 6 m^2 s N_C,
\ea
\ba &&
F_3^{(j)}=\frac{56}{3} \left\{
          2 \left[ 8 m^2 \left( \frac{1}{t} - \frac{z_2}{s^2\beta^2} \right)
                  - B_1^{(j)} \frac{2}{t} \left(m^2 + \frac{D}{s}\right)
                  - B_5^{(j)} \frac{2 z_u}{s\beta^2}     \right.\right.   \\ 
\nn && \left.  \qquad \qquad \qquad \qquad
             + C_1^{(j)} \left(\frac{4 t^2}{s} - z_2\frac{8 m^4 - s^2}{s^2\beta^2}\right)
             - C_3^{(j)} 2 t \left(1 + 2 \frac{T}{s}\right)
             + (C_4^{(j)} - D_2^{(j)} t) \frac{1}{s} (2 D + s^2 + 2 t^2)
                \right]                                          \\
\nn &&  \qquad \qquad
      + \ep \left[ - 8 m^2 \left( \frac{3}{t} - \frac{2 z_2}{s^2\beta^2} \right)
                  + B_1^{(j)} 2 \left( \frac{3 z_t}{t} + \frac{2 t}{s} \right)
                  - B_5^{(j)} 2 \left(2 + \frac{z_2}{s\beta^2}\right)    \right.    \\
\nn &&   \qquad \qquad \qquad \qquad
                  - C_1^{(j)} \left( 8 m^2 + 4 s + \frac{8 m^2 t + s^2}{s\beta^2}
                                  + 2 \frac{m^2 s^2 + 2 t^3}{D} \right)
                  + C_3^{(j)} 2 \frac{t}{s} \left(s - 4 t - 2 s t \frac{s-t}{D}\right) \\
\nn && \left.  \qquad \qquad \qquad \qquad
              - (C_4^{(j)} - D_2^{(j)} t) \left(3 s + 4 t + 2 s t \frac{s-t}{D} \right)
                  \right]                  \\
\nn && \left. \qquad \qquad
     + \ep^2 \frac{3 s^2}{D} \left[ C_1^{(j)} z_t + C_3^{(j)} \frac{2 t^2}{s} 
                        + C_4^{(j)} t - D_2^{(j)} t^2 \right]
\right\},
\ea
\ba  &&
F_4^{(j)}=\frac{16}{3} \left\{
          2 \left[ 8 m^2 \left( \frac{1}{u} - \frac{z_{2u}}{s^2\beta^2} \right)
                  - B_{1u}^{(j)} \frac{2}{u} \left(m^2 + \frac{D}{s}\right)
                  - B_5^{(j)} \frac{2 z_t}{s\beta^2}     \right.\right.   \\     
\nn && \left.  \qquad \qquad \qquad \qquad
           + C_1^{(j)} \left(\frac{4 u^2}{s} - z_{2u}\frac{8 m^4 - s^2}{s^2\beta^2}\right)
           - C_{3u}^{(j)} 2 u \left(1 + 2 \frac{U}{s}\right)
           + (C_4^{(j)} - D_{2u}^{(j)} u) \frac{1}{s} (2 D + s^2 + 2 u^2)  
                \right]                                          \\
\nn &&  \qquad \qquad
      + \ep \left[ - 8 m^2 \left( \frac{1}{u} + \frac{2 z_{2u}}{s^2\beta^2} \right)
               + B_{1u}^{(j)} 2 \left( \frac{z_u}{u} - \frac{2 u}{s} \right)
               - B_5^{(j)} 2 \left(1 - \frac{2 z_u}{s\beta^2}\right)    \right.    \\  
\nn &&   \qquad \qquad \qquad \qquad
               + C_1^{(j)} \left( 4 z_u + \frac{8 m^2 u + s^2}{s\beta^2}
                                  + 2 m^2 s \frac{z_{2u}}{D} \right)
               + C_{3u}^{(j)} 2 \frac{u}{s} \left(3 s + 4 u - 2 \frac{stu}{D}\right) \\
\nn && \left.  \qquad \qquad \qquad \qquad
              - (C_4^{(j)} - D_{2u}^{(j)} u) \left(2 \frac{m^2 s^2}{D} - s - 4 u \right)
                  \right]                  \\
\nn && \left. \qquad \qquad
     - \ep^2 \frac{3 s^2}{D} \left[ C_1^{(j)} z_u + C_{3u}^{(j)} \frac{2 u^2}{s}
                        + C_4^{(j)} u - D_{2u}^{(j)} u^2 \right]
\right\} .
\ea
\end{widetext}

The additional subscript ``u'' in some of the scalar coefficient functions in 
the expression for $F_4^{(j)}$ (such as $B_{1u}^{(j)}$) is to be understood 
as an operational definition 
prescribing a $(t\lra u)$ interchange in the argument of that function, i.e.
$B_{1u}^{(0)}=B_{1}^{(0)}\big|_{t\lra u}$, etc.
Note that $B_{5}^{j}$, $C_{1}^{j}$, and $C_{4}^{j}$ are intrinsically $(t\lra u)$ 
symmetric (see \cite{KMR}). 
Taking 
the $(t\lra u)$ symmetry of $B_{5}^{j}$, $C_{1}^{j}$, and $C_{4}^{j}$ into 
account, one notes a corresponding $(t\lra u)$ symmetry for the first and 
third square brackets in $F_3^{(j)}$ and 
$F_4^{(j)}$. 

Before presenting our result for the NLO matrix element, we would like to 
comment on its color 
structure. First note that all the vertex and self-energy (VSE) graphs are 
proportional to the LO Born term color matrices 
(see Refs. \cite{KM,KMR2}). Both the parallel ladder box, Fig.~\ref{fig:nlo}(a), 
and the crossed ladder box, Fig.~\ref{fig:nlo}(b), have their own color structures. 
Altogether one obtains the following three color structures,
\ba
\label{colnlo}
{\rm Tr} (T^a T^b) \,\, {\rm Tr} (T^a T^b) &\!=\!& \frac{d_A}{4}\Rightarrow 2,    \\
\nn
{\rm Tr} (T^a T^b T^c) \,\, {\rm Tr} (T^b T^a T^c)  &\!=\!&
  \frac{d_A}{8} \left( N_C - \frac{2}{N_C} \right) \Rightarrow \frac{7}{3}, \\
\nn
{\rm Tr} (T^a T^b T^c) \,\,{\rm Tr} (T^a T^b T^c) &\!=\!&
                 - \frac{d_A}{4} \frac{1}{N_C}  \Rightarrow - \frac{2}{3} ,
\ea
from folding the Born term with the VSE graphs, 
the parallel ladder box, Fig.~\ref{fig:nlo}(a), and the crossed ladder box, 
Fig.~\ref{fig:nlo}(b), in that order.
The common factor $d_A = N_C^2 - 1 = 8$ is the dimension of the adjoint 
representation of the color group ${\rm SU}(N_C)$. 
We present our NLO result separately for 
these three color structures.

At NLO the final spin and color summed matrix element can be written down as a 
sum of three terms:
\ba
\label{nloqq}
\nn   
\frac{1}{g_s^2 \sqrt\mathcal C}|M|^2_{{\rm Loop}\times{\rm Born}} &\!\!=\!\!&
                  {\rm Re}\Big[   \frac{1}{\ep^2} W^{(-2)}(\ep) +
                      \frac{1}{\ep} W^{(-1)}(\ep)        \\
&&
+ W^{(0)}(\ep) \Big],
\ea
where $\mathcal C$ has been defined in (\ref{common}). The notation
$|M|^2_{{\rm Loop}\times{\rm Born}}$ means that one is retaining only the
${\mathcal O} (\alpha_{s}^3)$ part of $|M|^{2}$.

The first two coefficient functions in (\ref{nloqq}) have a rather simple 
structure: 
\ba 
W^{(-2)}(\ep) &\!\!=\!\!& - 2 B (2 C_F - N_C + 3) ,    \\
\nn 
W^{(-1)}(\ep) &\!\!=\!\!&  - 2 B \Big( 5 C_F
         \Big[ C_4^{(-1)} s - C_6^{(-1)} (2 m^2 - s) \Big]    \\
\nn   & &\mbox{\hspn}
  \times  (2 C_F - N_C)
      - \frac{2}{3} \Big[ 7 \ln(\frac{-t}{m^2}) + 2 \ln(\frac{-u}{m^2})\Big]
\Big) \, ,
\ea
where $B$ is the Born term defined in Eq.~(\ref{lo}).
One should keep in mind that the overall Born term factor $B$ above contains a 
term multiplied by $\ep$. Therefore, if the expression for $B$, 
Eq.~(\ref{lo}), is substituted in $W^{(-2)}$ and $W^{(-1)}$, we will obtain 
$(\ep)^{-1}$ and finite terms from the first two terms of 
Eq.~(\ref{nloqq}).

The third term in Eq.~(\ref{nloqq}) reads 
\be
W^{(0)}(\ep) =  F_{\rm NLO}^{(0)} , 
\ee
where
\be
\label{nloterm}
F_{\rm NLO}^{(j)} = W_1^{(j)} + W_2^{(j)} + W_3^{(j)} ,
\ee
and where
\ba \nn  &&
\label{nloterms}
W_1^{(j)} = 2 B \,\, F_1^{(j)} + 128 \frac{m^2 D}{s^4\beta^4} F_2^{(j)} ,   \\
&&
W_2^{(j)} = - 2 B \beta_0 \ln^{1+j}(\frac{m^2}{\mu^2}) ,    \\
\nn    &&
W_3^{(j)} =  F_3^{(j)} + F_4^{(j)}.
\ea
Note that the first term in (\ref{nloterms}) originates entirely from the sum 
of self-energy and vertex diagrams while the second term is due to 
renormalization. The terms $F_3^{(j)}$ and $F_4^{(j)}$ in $W_3^{(j)}$ represent
the contributions from boxes $a$ and $b$, respectively.

The massless limit of our NLO result Eq.~(\ref{nloqq}) 
without the $\mathcal O(\ep)$ and $\mathcal O(\ep^2)$ order terms 
was compared (including also the imaginary part) with corresponding 
results obtained from the methods developed
in Ref.~\cite{MitovMoch} \footnote{The method of \cite{MitovMoch} can be 
considered to be a generalization of the results of
\cite{Catani} to higher orders in perturbation theory.}. There was agreement
\cite{Mitov}. This serves as a rigorous check on our singularity structure 
as well as on all the mass logarithms of our original NLO matrix element 
\cite{KM}. 



\section{\label{singular}
SINGULARITY STRUCTURE OF THE NNLO SQUARED AMPLITUDE
}


The NNLO final spin and color summed squared matrix element can be
written down as a sum of five terms:

\begin{eqnarray}
\label{nnlo}
\frac{1}{\mathcal C} |M|^2_{{\rm Loop}\times{\rm Loop}} &\!\!=\!\!&
                  {\rm Re}\Big[    \frac{1}{\ep^4} V^{(-4)}(\ep) +
                      \frac{1}{\ep^3} V^{(-3)}(\ep) \\
& &\mbox{\hspn}                    +\frac{1}{\ep^2} V^{(-2)}(\ep)      
                     + \frac{1}{\ep} V^{(-1)}(\ep) +
                                    V^{(0)}(\ep) \Big] , 
\nonumber
\end{eqnarray}
where $\mathcal C$ has been defined in (\ref{common}). Note Eq.~(\ref{nnlo}) is 
{\it not} a Laurent series expansion in $\ep$ since the coefficient functions
$V^{(m)}(\ep)$ are functions of $\ep$ as explicitly annotated in Eq.~(\ref{nnlo}).
It is nevertheless useful to write the NNLO one-loop squared result in the
form of Eq.~(\ref{nnlo}) in order to exhibit the explicit $\ep$ structures. 
All five coefficient functions $V^{(m)}(\ep)$
are bilinear forms in the coefficient functions that define the
Laurent series expansion of the scalar master integrals (\ref{Dexp}).
Some of these coefficient
functions are zero and some of them are just numbers or simple logarithms.
In the latter case we
will substitute these numbers or logarithms for the coefficient functions 
$V^{(m)}$ in the five terms above. This will be done for the coefficient 
functions 
$A^{(m)}$, $B_1^{(-1)}$,
$B_{1u}^{(-1)}$, $B_5^{(-1)}$, $C_3^{(-1)}$, and $C_{3u}^{(-1)}$.

%
We mention that in the course of our work 
we took full advantage of the fact that in \cite{KM} all the poles in the 
matrix element for the $q\bar{q}\ra Q{\overline Q}$ subprocess 
are multiplied only by the leading order Born Dirac structure 
to cast the singular terms of the squared matrix element into an appropriately 
factorized form.

Before proceeding further, we present three more color structures appearing in
the NNLO calculation in addition to the ones presented in Eq.~(\ref{colnlo}) :
\ba
\label{colnnlo}
{\rm Tr} (T^a T^b T^{b'} T^{a'})   \!\!\!&&\!\!\!
        {\rm Tr} (T^b T^a T^{a'} T^{b'}) =    \\
\nn    &&
      \frac{d_A}{16} \left[ N_C^2 - 3 + \frac{3}{N_C^2} \right] 
                     \Rightarrow  \frac{19}{6} ,   \\
\nn
{\rm Tr} (T^a T^b T^{b'} T^{a'})   \!\!\!&&\!\!\!
      {\rm Tr} (T^a T^b T^{b'} T^{a'})   =
         \frac{d_A}{16} \left[ 1 + \frac{3}{N_C^2} \right] 
                         \Rightarrow  \frac{2}{3} , \\
\nn
{\rm Tr} (T^a T^b T^{b'} T^{a'})   \!\!\!&&\!\!\!
       {\rm Tr} (T^b T^a T^{b'} T^{a'}) =       \\
\nn   &&
              - \frac{d_A}{16}\left[ 1 - \frac{3}{N_C^2}  \right] 
                                         \Rightarrow   - \frac{1}{3} .
\ea
The above three color structures arise from folding 
box $a$ with box $a$, 
box $b$ with box $b$, as well as the interference of the two boxes, respectively.

Let us first introduce a notation which will help us to present the coefficients of
the singular terms in the most concise fashion:
\ba
\label{comlo}
\nn  &&
L_{1} = (2 C_F - N_C) \left(C_4^{(-1)} s - C_6^{(-1)} (2 m^2 - s)\right),  \\
\nn  &&
L_{2} = 15 C_F - 14 \ln(\frac{-t}{m^2}) - 4 \ln(\frac{-u}{m^2}),    \\
\nn  &&
L_{3} = 35 C_F - 38 \ln(\frac{-t}{m^2}) - 4 \ln(\frac{-u}{m^2}),    \\
&&
L_{4} = 5 C_F - 2 \ln(\frac{-t}{m^2}) - 4 \ln(\frac{-u}{m^2}).
\ea

The two most singular terms in (\ref{nnlo}) are proportional 
to the Born term $B$ defined in (\ref{lo}). One has 
\ba
\label{eps43}
V^{(-4)}(\ep) &\!=\!& (2 C_F - N_C + 3)^2 B ,    \\
\nn
V^{(-3)}(\ep) &\!=\!&  2 (2 C_F - N_C + 3) B 
              \left[ L_{1} + \frac{L_{2}}{3}\right].
\ea
We also obtain
\ba
\label{eps2}
V^{(-2)}(\ep) &\!=\!& \frac{B}{3} \Big[ 
       (3 L_{1} + L_{2}) (L_{1} + 5 \cf)^\ast    \\
\nn  & &
       - 2 \ln(\frac{-t}{m^2}) ( 7 L_{1} + L_{3} )  \\
\nn &&
       - 4 \ln(\frac{-u}{m^2}) ( L_{1} + L_{4} )
\Big]             \\
\nn  &&
- (2\*\cf - \nc + 3) F_{\rm NLO}^{(0)} .
\ea
The last term in Eq.~(\ref{eps2}) is obtained from folding the 
${\mathcal O}(\ep^{-2})$ singular term of the matrix element with its finite part, 
while the rest is obtained from folding the single poles.
Note that when one substitutes the Laurent expansions for $B$ and $F_{\rm NLO}^{(0)}$, 
one gets additional $1/\ep$ poles and finite terms in Eq.~(\ref{eps2}).

The structure of the last term in Eq.~(\ref{nnlo}) is a little more complicated. One has
\ba
\label{eps}
V^{(-1)}(\ep) &\!\!=\!\!& - L_{1}^\ast F_{\rm NLO}^{(0)} 
                          - \frac{L_{2}}{3} ( W_1^{(0)} + W_2^{(0)} )    \\
\nn  &&
                          - \frac{L_{3}}{7} F_3^{(0)} - L_{4} F_4^{(0)}  \\
\nn &&
 + (2 C_F - N_C + 3) \left[ - F_{\rm NLO}^{(1)} + V^{'} \right] .
\ea
The terms multiplied by the $L_{m}$ functions above are due to 
folding the single pole terms in the original matrix element  
with its finite $\mathcal O(\ep^0)$ part, while the last term is due
to interference $\mathcal O(\ep^{-2})\times \mathcal O(\ep)$ terms in 
the original matrix element. This pole term is due to the Taylor 
expansion of the original matrix element and cannot be
deduced from the knowledge of the LO terms alone.
The function $F_{\rm NLO}^{(1)}$ is defined in Eq.~(\ref{nloterm}) and is
nothing but the finite part of the 
NLO term with indices of the
coefficient functions of the scalar master integrals and the power of 
the logarithm that multiplies the $\beta_0$ function, shifted upward by 1. 
For the remaining term $V^{'}$, one obtains 

\begin{widetext}
\ba
&&
V^{'} = - 2 B \left[ \frac{\beta_0}{2} \ln^2(\frac{m^2}{\mu^2})
            + 8 C_F - \frac{N_C}{\beta^2} - \frac{2 n_{l} + 2 + 28 N_C}{27}
            + B_2^{(0)} \frac{2\beta^2 - 18 C_F + 9 N_C}{9}    \right.     \\
\nn  &&   \left.   \qquad
            + B_5^{(0)} \frac{2}{9} (5 N_C + n_{l} - 9 C_F) \right]
      - 128 \frac{m^2 D}{s^3\beta^4} \left[
              2 (6 \beta^2 C_F - N_C)
            - B_2^{(0)} 2 \beta^2 (2 C_F - N_C)
            - B_5^{(0)} 2 N_C - C_1^{(0)} s N_C \right]       \\
\nn  &&
    - \frac{56}{3} \left\{ 2 \left[ 8 m^2 \left(\frac{1}{t} - \frac{z_2}{s^2\beta^2}\right)
            + \left(\frac{2}{s} + \frac{s - t}{D}\right)
                       \left( C_1^{(0)} s z_t + C_3^{(0)} 2 t^2
                            + C_4^{(0)} s t - D_2^{(0)} s t^2 \right) \right]
\right.  \\
\nn   &&    \qquad   \left.
         - \ep \left[
             8 m^2 \left(\frac{3}{t} - \frac{2 z_2}{s^2\beta^2}\right)
          +  \left(\frac{8}{s} + \frac{7 s - 4 t}{D}\right)
                     \left( C_1^{(0)} s z_t + C_3^{(0)} 2 t^2
                          + C_4^{(0)} s t - D_2^{(0)} s t^2 \right) \right]
\right\}
\\
\nn   &&
    - \frac{16}{3} \left\{ 2 \left[ 8 m^2 \frac{z_{2u}}{s^2\beta^2}
            + B_{1u}^{(0)} 2 \left( \frac{2 D}{s u} - 1\right)
            - B_5^{(0)} \frac{2 z_{2u}}{s\beta^2}
            - C_1^{(0)} \left( m^2 \left(4 + \frac{s z_{2u}}{D}\right)
                                               - \frac{2 z_t}{\beta^2}\right)
\right. \right.  \\
\nn  &&   \left.     \qquad
            - \left( \frac{z_{2u}}{s} - \frac{t u}{D}\right)
        \left( C_{3u}^{(0)} 2 u + C_4^{(0)} s - D_{2u}^{(0)} s u\right) \right]   \\
\nn  &&   \qquad       \left.
         + \ep \left[
            - 8 m^2 \left( \frac{1}{u} + \frac{2 z_{2u}}{s^2\beta^2} \right)
            + \left( \frac{8}{s} + \frac{9 s + 4 u}{D} \right)
            \left( C_1^{(0)} s z_u + C_{3u}^{(0)} 2 u^2
            + C_4^{(0)} s u - D_{2u}^{(0)} s u^2 \right) \right]
\right\}.
\ea
\end{widetext}

When one substitutes the Laurent expansions for $F_3^{(0)}$, $F_4^{(0)}$, and
$F_{\rm NLO}^{(1)}$, one gets finite and ${\mathcal O}(\ep)$ order 
terms in Eq.~(\ref{eps}). 
However, since
we are only interested in the Laurent series expansion up to the finite
term, these $\mathcal O(\ep)$ contributions should be omitted.


\section{\label{finite}
STRUCTURE OF THE FINITE PART
}

In this section we present the finite part of our result. We calculate the 
finite part in several pieces, e.g.
\be
\label{fin}
V^{(0)} = {\rm Re}\left[V_{Bf_1}^{(0)} + V_{Bf_2}^{(0)}+ V_{f_0f_0}^{(0)}\right].
\ee

The first two terms originate from the interference of the 
$\mathcal O(\ep^{-1})\times \mathcal O(\ep)$ and 
$\mathcal O(\ep^{-2})\times \mathcal O(\ep^2)$ pieces of the initial
matrix element.
Each of them can be conveniently presented as a sum of five 
compact expressions:
\be
V_{Bf_1}^{(0)} = G_{1} + G_{2} + G_{3} + G_{4} + G_{5},
\ee
where
\ba
\label{eq:fb11}
  G_{1} &\! =\! &
         - 128 \* m^2 \* D \* ( L_{1}^\ast + L_{2}/3 ) \*
           \Big[ F_2^{(1)} 
\nonumber
\\
& &\mbox{\hspn}
          + 12 \* s \* \beta^2 \* \cf - 2 \* s \* \nc
 - B_{2}^{(0)} \* 2 \* s \* \beta^2 \* (2 \* \cf - \nc)
\nn  \\
& &\mbox{\hspn}
- B_{5}^{(0)} \* 2 \* s \* \nc
- C_{1}^{(0)} \* s^2  \* \nc  \Big]/(s^4 \* \beta^4)
\nonumber
\; ,
\quad
\\[2mm]
\label{eq:fb12}
  G_{2} &\! =\! &
 - 2 \* B \* ( L_{1}^\ast + L_{2}/3 ) \*
              \Big[ 27 \* F_1^{(1)}
              - 2 \* \nl - 2          
\nonumber\\
& &\mbox{\hspn}
- 28 \* \nc + 216 \* \cf - 27 \* \nc/\beta^2
              - B_{2}^{(0)} \* 3 \* (18 \* \cf 
\nonumber\\
& &\mbox{\hspn}
- 9 \* \nc - 2 \* \beta^2)
              - B_{5}^{(0)} \* 6 \* (9 \* \cf - 5 \* \nc - \nl) \Big]/27
\nonumber
\; ,
\quad
\\[2mm]
\label{eq:fb13}
  G_{3} &\! =\! &
 \beta_0 \* B \* \ln^2(\frac{m^2}{\mu^2}) \* ( L_{1} + L_{2}/3 )
\; ,
\quad
\\[2mm]  
\label{eq:fb14}
  G_{4} &\! =\! &
    - 16 \* ( 7 \* L_{1}^\ast + L_{3} ) \*
     \Big[ F_3^{(1)}\, \* 3/112  
\nonumber \\ 
& &\mbox{\hspn}
     + 8 \* m^2 \* (1/t - z_2  /(s^2 \beta^2) )
 + (C_{1}^{(0)} \* \zt + C_{3}^{(0)} \* 2 \* t^2/s 
\nonumber \\
& &\mbox{\hspn}
+ C_{4}^{(0)} \* t - D_{2}^{(0)} \* t^2) \*
(2 \* D + s^2 - s \* t)/D \Big]/3
\nonumber
\; ,
\quad
\\[2mm]
\label{eq:fb15}
  G_{5} &\! =\! &
 - 32 \* ( L_{1}^\ast + L_{4} ) \*
     \Big[ F_4^{(1)}\, \* 3/32 
\nonumber \\
& &\mbox{\hspn}
           + 8 \* m^2 \* z_{2u}/(s^2 \beta^2)
           + B_{1u}^{(0)} \* 2 \* (2 \* D/(s u) - 1)
\nonumber \\
& &\mbox{\hspn}
- B_{5}^{(0)} \* 2 \* z_{2u}/(s \beta^2)
           - C_{1}^{(0)} \* (m^2 \* s \* z_{2u}/D 
\nonumber \\
& &\mbox{\hspn}
- 2 \* (8 \* m^4 + s \* t)/(s \beta^2))
           - (C_{3u}^{(0)} \* 2 \* u/s + C_{4}^{(0)} 
\nonumber \\
& &\mbox{\hspn}
- D_{2u}^{(0)} \* u) \* (z_{2u} - s \* t
\* u/D) \Big]/3
\nonumber
\; .
\ea

The first three terms above are due to the VSE 
contributions, and the last two terms 
are due to the two box diagrams. 
Similarly, for the second term in Eq.~(\ref{fin}) we write
\begin{widetext}
\be
V_{Bf_2}^{(0)} = H_{1} + H_{2} + H_{3} + H_{4} + H_{5},
\ee
with
\begin{eqnarray}
\label{eq:BF211fin}
  H_{1} &\! =\! &
 - 128 \* (2 \* \cf - \nc + 3) \* D \* m^2 \* \Big[
              F_2^{(2)}
            + 4 \* s \* \beta^2 \* (7 \* \cf + \nc) - 10 \* s \* \nc  
            - B_{2}^{(1)} \* 2 \* s \* \beta^2 \* (2 \* \cf - \nc)
\nn \\
& &\mbox{\hspn}
            - B_{5}^{(1)} \* 2 \* s \* \nc
            - C_{1}^{(1)} \* s^2 \* \nc \Big]/(s^4 \* \beta^4)
\nonumber
\; ,
\quad
\\[2mm]
\label{eq:BF212fin} 
  H_{2} &\! =\! &
 - (2 \* \cf - \nc + 3) \* B \* \Big[
              \* F_{1}^{(2)} \* 162
            + 2 \* (1296 \* \cf + 76 \* \nc - 10 \* \nl - 10 - 243 \* \nc/\beta^2)
            + B_{2}^{(0)} \* 24 \* \beta^2
\nonumber \\
& &\mbox{\hspn}
            - B_{2}^{(1)} \* 18 \* (18 \* \cf - 9 \* \nc - 2 \* \beta^2)
            + B_{5}^{(0)} \* 12 \* (\nc + 2 \* \nl)
            - B_{5}^{(1)} \* 36 \* (9 \* \cf - 5 \* \nc - \nl) \Big]/81
\nonumber
\; ,
\quad
\\[2mm]
\label{eq:BF213fin}
 H_{3}  &\! =\! &
 (2 \* \cf - \nc + 3) \* B \* \beta_0 \* \ln^3(\frac{m^2}{\mu^2})/3
\; ,
\quad
\\[2mm]
\label{eq:BF22fin}
 H_{4}  &\! =\! & 
 - 112 \* (2 \* \cf - \nc + 3) \* \Big[
        F_3^{(2)}\, \* 3/112
     + 24 \* m^2 \* (1/t - z_2 /(s^2\beta^2))
     + ( \zt \* (2 \* C_{1}^{(0)} + C_{1}^{(1)})
\nonumber \\
& &\mbox{\hspn}
 + 2 \* t^2 \* (2 \* C_{3}^{(0)} + C_{3}^{(1)})/s
       + t \* (2 \* C_{4}^{(0)} + C_{4}^{(1)}) - t^2 \* (2 \* D_{2}^{(0)} + D_{2}^{(1)}) ) \*
(2 \* D + s^2 - s \* t)/D \Big]/3
\nonumber
\; ,
\quad
\\[2mm]
\label{eq:BF23fin}
  H_{5} &\! =\! &
 - 32 \* (2 \* \cf - \nc + 3) \* \Big[
        F_4^{(2)}\, \* 3/32
     + 8 \* m^2 \* (1/u + z_{2u}/(s^2\beta^2))
     + B_{1u}^{(1)} \* 2 \* (2 \* D/(s u) - 1)
\nonumber \\
& &\mbox{\hspn}
     - B_{5}^{(1)} \* 2 \* z_{2u}/(s \beta^2)
     - ( C_{1}^{(0)} \* \zu + C_{3u}^{(0)} \* 2 \* u^2/s + C_{4}^{(0)} \* u - D_{2u}^{(0)} \*
u^2 ) \* (4 \* D + 3 \* s^2 + 2 \* s \* u)/D
\nonumber \\
& &\mbox{\hspn}
     - C_{1}^{(1)} \* (m^2 \* s \* z_{2u}/D - 2 \* (8 \* m^4 + s \* t)/(s \beta^2))
     - ( C_{3u}^{(1)} \* 2 \* u/s + C_{4}^{(1)} - D_{2u}^{(1)} \* u ) \* (z_{2u} \* D - s \* t
\* u)/D \Big]/3
\nonumber
\; .
\end{eqnarray}
\end{widetext}

Note again that the $\mathcal O(\ep)$ and $\mathcal O(\ep^2)$ 
order terms in the above expressions 
for $V_{Bf_1}^{(0)}$ and $V_{Bf_2}^{(0)}$ can be disregarded. 
We also mention that
the scalar coefficient functions with superscript ``2'' above involve
multiple polylogarithms.

We emphasize that the factorized forms of all the expressions given in 
this paper hold only when one retains the full $\ep$ dependence in the Born
and NLO terms.

The last term in Eq.~(\ref{fin}) comes from the square of the $O(\ep^0)$ 
term of the matrix element. It can also be written as a sum of five
terms:
\be
\label{factors}
V_{f_0f_0}^{(0)} = M_{VSE} + M_{BVSE} + M_{aa} + M_{ba} + M_{bb}.
\ee

The first term is the square of the finite parts of vertex and self-energy
graphs; the second one is an interference of the vertex and self-energy
graphs with the two box diagrams. These two terms can be presented in 
a very compact form:
\ba  \nn
M_{\rm VSE}&\!=\!& F_1^{(0)} \left( W_1^{(0)} + W_2^{(0)} - B \,F_1^{(0)} \right)^\ast  \\
 && \mbox{\hspn}
             - \big|F_2^{(0)}\big|^2 \, 32 m^2 D/(s^5\beta^6)   \\
\nn  && \mbox{\hspn}
  - \beta_0 \ln(\frac{m^2}{\mu^2}) 
                        \left( W_2^{(0)}/2 + W_1^{(0)} - 2B\, F_1^{(0)} \right);   \\
\nn
M_{BVSE} &\!=\!& 7 P + 2 P|_{t\lra u}
      + \left( F_1^{(0)} - \beta_0 \ln(\frac{m^2}{\mu^2})\right)  \\
  && \mbox{\hspn}
                    \times ( F_3^{(0)} + F_4^{(0)})^\ast ,
\ea
with 
\ba   \nn
P &\!\!=\!\!& 64 m^2 F_2^{(0)\ast} \, \Big[
              2 D/t - B_{1}^{(0)} D/t + C_1^{(0)} T z_2 - C_{3}^{(0)} 2 t T \\
  && \mbox{\hspn}
            + (C_4^{(0)} - D_2^{(0)} t) (D + t^2) \Big]/(3 s^3 \beta^4) .
\ea
When writing out $P|_{t\lra u}$ one has to do the $t\lra u$ operation in all the 
terms in the function $P$, i.e. for $z,t,F_{2}^{(0)},B_{1}^{(0)},C_{3}^{(0)},T$, and 
$D_{2}^{(0)}$ ($C_{1}^{(0)}$ and $C_{4}^{(0)}$ are $t\lra u$ symmetric).

Finally, we are left with the last three terms in Eq.~(\ref{factors}), which 
are the longest terms in our NNLO result. However, to our surprise, 
we were able to discover nice factorization properties of 
the square of the two box diagrams. This part of the cross section can be 
put together with the help of several  
building blocks; e.g. each of the last three terms in Eq.~(\ref{factors}) 
can be written as a sum of bilinear products. Each of the factors in the
bilinear products are linear
combinations of scalar integral coefficient functions multiplied by
some combination of kinematic variables. To be more specific, we write
\ba 
\label{boxes}
\nn
M_{aa} &\!=\!& \frac{76}{3} \Big[ s^{-1} Q_{1} Q_{8}^\ast 
                                 + 4 m^2 Q_{2} Q_{3}^\ast
                                         + Q_{4} Q_{10}^\ast         \\
\nn  &&  \mbox{\hspn}
               + m^2 Q_{5} Q_{11}^\ast
               - 2 s^{-1} Q_{6} Q_{12}^\ast + Q_{7} Q_{13}^\ast  \Big],   \\
M_{bb} &\!=\!& \frac{4}{19} M_{aa}|_{t\lra u} ,        \\
\nn
M_{ba} &\!=\!& \frac{16}{3} \Big[ s^{-1} Q_{8} Q_{14}^\ast 
                                   + 4 m^2 Q_{9} Q_{15}^\ast
                                         + Q_{10} Q_{16}^\ast         \\
\nn  &&  \mbox{\hspn}
               + 2 m^2 Q_{11} Q_{16}^\ast
               + 2 s^{-1} Q_{12} Q_{17}^\ast + Q_{13} Q_{18}^\ast  \Big].
\ea
Explicit expressions for the 18 linear forms $Q_{i}$ are given 
in the Appendix. The bilinear forms above arise from folding 
certain pairs of Dirac structures in our original matrix element.
The expression for $M_{ba}$ represents the result of the interference
of the finite parts of box $a$ and box $b$. 

It is quite obvious that the factorized forms for the finite part of
the NNLO result for the $q\bar{q}\ra Q\overline{Q}$ subprocess should also hold 
for the corresponding massless amplitudes. We have not seen this being
displayed in the literature.

In the finite contribution, Eq.~(\ref{fin}), one can see the interplay of the
product of powers of $\ep$ resulting from the Laurent series expansion of the
scalar integrals [cf.~ Equation~(\ref{Dexp})] on the one hand, and powers of $\ep$
resulting from doing the spin algebra in dimensional regularization on the other 
hand. For example, for the finite part
one has a contribution from $C_6^{(-1)}B_1^{(0)\ast}$ as well as a contribution from
$C_6^{(-1)}B_1^{(1)\ast}$. Terms of the type $C_6^{(-1)}B_1^{(0)\ast}$, where the
superscripts corresponding to
$\ep$ powers do not compensate, would be absent in the regularization schemes
where traces are effectively taken in four dimensions, i.e. in the so-called
four-dimensional schemes or in dimensional reduction.

We want to emphasize that all our factorized 
results given in this paper take up about
10 Kb of hard disk space. This has to be compared with the length of the 
original computer output.
The original computer output for the one-loop squared 
cross section of the $q\bar q\ra Q{\overline Q}$ subprocess
turned out to be very long and took up approximately 4~MB of hard disk space. 
Therefore, the reduction is of the order of 400 in the present case.

As a final remark we want to emphasize that we have done two independent 
calculations using REDUCE \cite{reduce} and FORM \cite{form} when squaring
the one--loop amplitudes. 
After casting the results into the above
compact
form, we have checked the final result against the original untreated versions 
using again the REDUCE Computer Algebra System.


\section{\label{summary}
CONCLUSIONS
}


We have presented analytical ${\cal O}(\alpha_s^4)$ NNLO results
for the one-loop squared contributions to heavy quark pair production in
quark--antiquark annihilation. These are the first exact results for the 
hadroproduction of heavy quarks at NNLO where the heavy quark mass dependence 
is fully retained. Our results form part of the NNLO description of heavy quark
pair production relevant for the NNLO analysis of ongoing experiments at the 
Tevatron and planned experiments at the LHC.

Our analytical results are presented in factorized forms. For the divergent 
parts, the squared matrix elements are given in terms of the Laurent series 
expansion of the corresponding LO and NLO contributions expanded up to
${\cal O}(\ep)$ and ${\cal O}(\ep^2)$, respectively.
In this way we could transfer parts of the finite part of the squared 
amplitudes to the coefficient functions of the pole terms. 
After this we found that the remaining parts of the finite contribution 
could be further factorized, partly in terms of the corresponding LO and NLO 
pieces, and, for the box graphs, in terms of factorizing forms as described 
in Sec.~\ref{finite}. Writing our analytical results in factorized forms 
led to a reduction of the length of the original output by a factor of 400. 
To the best of our knowledge these nice factorization
properties of the squared amplitude were not known before. It would be 
interesting to find out the underlying reason for this factorization. 

The present paper deals with unpolarized quarks in the initial and final states.
Since our results for the matrix elements in \cite{KMR2} contain the full spin 
information of the process, an extension to the polarized case with 
polarization in the initial state and/or in the final state including spin 
correlations is not difficult.

The present calculation constitutes a first step in obtaining the full NNLO 
corrections to the heavy quark production processes in QCD. A further next 
step is to provide one--loop squared results for gluon-initiated heavy quark 
pair production. Work on the gluon--initiated channel is in progress.

Analytical results in electronic format for all the terms in Eq.~(\ref{nnlo}),
including the $(t\lra u)$ symmetric terms explicitly written out, as well as 
combined full results, are readily available
\footnote{All the relevant results are available in {\sc Reduce} and {\sc Form} format. 
          The results
          can be retrieved from the preprint server 
          {\tt http://arXiv.org} by downloading the source of
          this article or can be obtained directly from the authors.}.

\begin{acknowledgments}
We would like to thank J.~Gegelia and B.~Kniehl for 
useful discussions. 
Z.M. acknowledges a very helpful communication with A.~Grozin.
Z.M. would like to thank the Particle Theory Group of the
Institut f{\"u}r Physik, Universit{\"a}t Mainz for hospitality, where large
parts of this work were done.
The work of Z.M. was supported by the German Research Foundation DFG 
through Grant No. KO~1069/11-1, 
in part by the Georgia National Science Foundation through Grant No.  
GNSF/ST07/4-196, by the German Federal Ministry for Education
and Research BMBF through Grant No.~05 HT6GUA, and by the DFG through Grant 
No.~KN~365/7-1.
M.R. was supported by the Helmholtz Gemeinschaft
under Contract No. VH-NG-105.
\end{acknowledgments}

\begin{widetext}
\appendix*
\section{}

Here we present the expressions for the terms $Q_{m}$ appearing in 
Equation (\ref{boxes}).

\begin{eqnarray}
\label{eq:QAA1}
  Q_{1} &\! =\! &
         \Big[ 8 \* m^2 \* (s/t - z_2 /(s \* \beta^2))
              - B_{1}^{(0)} \* 2 \* (m^2 \* s + D)/t
              - B_{5}^{(0)} \* 2 \* \zu/\beta^2
              + C_{1}^{(0)} \* (2 \* D + s^2 + 2 \* t \* \zt + 2 \* m^2 \* z_2 /\beta^2 )
\nonumber \\
& &\mbox{\hspn}
              - C_{3}^{(0)} \* 2 \* t \* (s + 2 \* T)
              + C_{4}^{(0)} \* (2 \* D + s^2 + 2 \* t^2)
              - D_{2}^{(0)} \* t \* (2 \* D + s^2 + 2 \* t^2) \Big]/D
\nonumber
\; ,
\quad
\\[2mm]
\label{eq:QAA2}
  Q_{2} &\! =\! &
          2/t - B_{1}^{(0)}/t + C_{1}^{(0)} \* T \* z_2 /D - C_{3}^{(0)} \* 2 \* t \* T/D
              + C_{4}^{(0)} \* (1 + t^2/D) - D_{2}^{(0)} \* t \* (1 + t^2/D)
\; ,
\quad
\\[2mm]
\label{eq:QAA3}
  Q_{3} &\! =\! &
          4 \* (2 \* m^2 \* \zt - D)/(s \* \beta^2 \* t)
              + B_{1}^{(0)} \* 2 \* T/t
              + B_{5}^{(0)} \* 2 \* \zt/(s \* \beta^2 )
              + C_{1}^{(0)} \* (\zt/\beta^2 + (m^2 \* s \* t - t^3)/D)
\nonumber \\
& &\mbox{\hspn}
              + C_{3}^{(0)} \* 2 \* t^3/D
              + C_{4}^{(0)} \* s \* t^2/D
              - D_{2}^{(0)} \* s \* t^3/D
\nonumber
\; ,
\quad
\\[2mm]
\label{eq:QAA4}
  Q_{4} &\! =\! &
         \Big[ 8 \* m^2 \* \zu/(s \* \beta^2)
               + B_{1}^{(0)} \* 2 \* m^2
               + B_{5}^{(0)} \* 2 \* m^2 \* z_2 /(s \* \beta^2)
               + C_{1}^{(0)} \* (2 \* t \* T + m^2 \* z_2 /\beta^2)
\nonumber \\
& &\mbox{\hspn}
               - C_{3}^{(0)} \* 2 \* m^2 \* t
               + C_{4}^{(0)} \* (m^2 \* s + 2 \* t^2)
               - D_{2}^{(0)} \* t \* (m^2 \* s + 2 \* t^2) \Big]/D
\nonumber
\; ,
\quad  
\\[2mm]
\label{eq:QAA5}
  Q_{5} &\! =\! &
         \Big[ 16 \* m^2 \* \zu/(s \* \beta^2)
               + B_{1}^{(0)} \* 4 \* m^2
               + B_{5}^{(0)} \* 4 \* m^2 \* z_2 /(s \* \beta^2)
               + C_{1}^{(0)} \* 2 \* (2 \* t \* T + m^2 \* z_2 /\beta^2)
\nonumber \\
& &\mbox{\hspn}
               - C_{3}^{(0)} \* 4 \* m^2 \* t
               + C_{4}^{(0)} \* 2 \* (m^2 \* s + 2 \* t^2)
               - D_{2}^{(0)} \* 2 \* t \* (m^2 \* s + 2 \* t^2) \Big]/D
\nonumber
\; ,
\quad  
\\[2mm]
\label{eq:QAA6}
  Q_{6} &\! =\! &
          16 \* m^2/(s \* \beta^2)            
              + B_{1}^{(0)} \* 2
              - B_{5}^{(0)} \* 2/\beta^2
              - C_{1}^{(0)} \* (4 \* t \* (D + m^2 \* t) + s^2 \* T/\beta^2 + 4 \* m^2 \*
t^2/\beta^2)/D
\nonumber \\
& &\mbox{\hspn}
              + C_{3}^{(0)} \* 2 \* t \* T \* z_2 /D
              - C_{4}^{(0)} \* z_2  \* (D + t^2)/D
              + D_{2}^{(0)} \* t \* z_2  \* (D + t^2)/D
\nonumber
\; ,   
\quad
\\[2mm]
\label{eq:QAA7}
  Q_{7} &\! =\! &
        \Big[  8 \* m^2 \* (s/t - 4 - 5 \* \zt/(s \* \beta^2))
               - B_{1}^{(0)} \* 2 \* (2 \* D/t - 3 \* m^2 + u)
               + B_{5}^{(0)} \* 2 \* (m^2 \* s + 6 \* m^2 \* t - s \* u)/(s \* \beta^2)
\nonumber \\
& &\mbox{\hspn}
               + C_{1}^{(0)} \* (2 \* m^2 \* s + 10 \* t \* T + (m^2 + s) \* z_2 /\beta^2)
               - C_{3}^{(0)} \* 2 \* t \* (5 \* m^2 + z_2 )
               + C_{4}^{(0)} \* (s \* (5 \* m^2 + z_2 ) + 10 \* t^2)
\nonumber \\
& &\mbox{\hspn}
               - D_{2}^{(0)} \* t \* (s \* (5 \* m^2 + z_2 ) + 10 \* t^2) \Big]/D
\nonumber
\; ,
\quad
\\[2mm]
\label{eq:QA1}
  Q_{8} &\! =\! &
            8 \* m^2 \* (D/t - t \* z_2 /(s \* \beta^2))
               - B_{1}^{(0)} \* 2 \* T \* (2 \* D/t - s)
               + B_{5}^{(0)} \* 2 \* (D + t \* \zt/\beta^2)
\nonumber \\
& &\mbox{\hspn}
               - C_{1}^{(0)} \* s \* (m^2 \* s - t^2 - t \* \zt - t \* \zt/\beta^2
                                    - t^2 \* (m^2 \* s - t^2)/D)
               - C_{3}^{(0)} \* 2 \* s \* t \* T \* (1 + s \* t/D)
\nonumber \\
& &\mbox{\hspn}
               + C_{4}^{(0)} \* s \* (m^2 \* s + 2 \* t^2 + s \* t^3/D)
               - D_{2}^{(0)} \* s \* t \* (m^2 \* s + 2 \* t^2 + s \* t^3/D)
\nonumber
\; ,
\quad
\\[2mm]
\label{eq:QA2}
  Q_{9} &\! =\! &
           - 4 \* (T/t + \zt/(s \* \beta^2))
                + B_{1}^{(0)} \* 2 \* T/t
                + B_{5}^{(0)} \* 2 \* \zt/(s \* \beta^2)
                + C_{1}^{(0)} \* (\zt/\beta^2 + t \* (m^2 \* s-t^2)/D)
                + C_{3}^{(0)} \* 2 \* t^3/D
\nonumber \\
& &\mbox{\hspn}
                + C_{4}^{(0)} \* s \* t^2/D
                - D_{2}^{(0)} \* s \* t^3/D
\nonumber
\; ,
\quad
\\[2mm]
\label{eq:QA3}
  Q_{10} &\! =\! &
           \Big[ 8 \* m^2 \* D
                - B_{1}^{(0)} \* 2 \* D \* \zt
                + B_{5}^{(0)} \* 2 \* t \* D
                - C_{1}^{(0)} \* s \* t \* (m^2 \* s + 4 \* m^2 \* t + t^2)
                + C_{3}^{(0)} \* 2 \* t^2 \* (m^2 \* s - t^2)
 \nonumber \\
& &\mbox{\hspn}
                + C_{4}^{(0)} \* s \* t \* (m^2 \* s - t^2)
                - D_{2}^{(0)} \* s \* t^2 \* (m^2 \* s - t^2) \Big]/D
\nonumber
\; ,
\quad  
\\[2mm]
\label{eq:QA4}
  Q_{11} &\! =\! &
           \Big[ 8 \* D \* \zu/(s \* \beta^2)
                + B_{1}^{(0)} \* 2 \* D
                + B_{5}^{(0)} \* 2 \* z_2  \* D/(s \* \beta^2)
                - C_{1}^{(0)} \* s \* (m^2 \* s - t^2 - z_2  \* D/(s \* \beta^2))
                - C_{3}^{(0)} \* 2 \* s \* t^2
\nonumber \\
& &\mbox{\hspn}
                - C_{4}^{(0)} \* s^2 \* t
                + D_{2}^{(0)} \* s^2 \* t^2 \Big]/D
\nonumber
\; ,
\quad  
\\[2mm]
\label{eq:QA5}
  Q_{12} &\! =\! &
           8 \* m^2 \* \zt/(s \* \beta^2)
              + B_{1}^{(0)} \* 2 \* T
              - B_{5}^{(0)} \* 2 \* (D - t \* \zt)/(s \* \beta^2)
              + C_{1}^{(0)} \* s \* \zt \* ((2 \* m^2 - s)/(s\beta^2) + t^2/D)
\nonumber \\
& &\mbox{\hspn}
              - C_{3}^{(0)} \* 2 \* s \* t^2 \* T/D
              - C_{4}^{(0)} \* s^2 \* t \* T/D
              + D_{2}^{(0)} \* s^2 \* t^2 \* T/D
\nonumber
\; ,
\quad  
\\[2mm]
\label{eq:QA6}
  Q_{13} &\! =\! &
           C_{1}^{(0)} \* s \* \zt + C_{3}^{(0)} \* 2 \* t^2 + C_{4}^{(0)} \* s \* t -
D_{2}^{(0)} \* s \* t^2
\nonumber
\; ,
\quad
\\[2mm]
\label{eq:QB1}
  Q_{14} &\! =\! &
 \Big[ 8 \* m^2 \* (s/u - z_{2u}/(s \* \beta^2))
                - B_{1u}^{(0)} \* 2 \* (m^2 \* s + D)/u
                - B_{5}^{(0)} \* 2 \* \zt/\beta^2
                + C_{1}^{(0)} \* (2 \* D + s^2 + 2 \* u \* \zu + 2 \* m^2 \* z_{2u}/\beta^2)
\nonumber \\
& &\mbox{\hspn}
                - C_{3u}^{(0)} \* 2 \* u \* (s + 2 \* U)
                + C_{4}^{(0)} \* (2 \* D + s^2 + 2 \* u^2)
                - D_{2u}^{(0)} \* u \* (2 \* D + s^2 + 2 \* u^2) \Big]/D
\nonumber
\; ,
\quad
\\[2mm]
\label{eq:QB2}
  Q_{15} &\! =\! &
 2/u - B_{1u}^{(0)}/u + C_{1}^{(0)} \* U \* z_{2u}/D - C_{3u}^{(0)} \* 2 \* u \* U/D
              + C_{4}^{(0)} \* (1 + u^2/D) - D_{2u}^{(0)} \* u \* (1 + u^2/D)
\nonumber
\; ,
\quad
\\[2mm]
\label{eq:QB3}
  Q_{16} &\! =\! &
           \Big[ 8 \* m^2 \* (s/u - \zu/(s \* \beta^2))
                - B_{1u}^{(0)} \* 2 \* (D - m^2 \* t)/u
                - B_{5}^{(0)} \* 2 \* (t - m^2 \* z_{2u}/(s \* \beta^2))
\nonumber \\
& &\mbox{\hspn}
                + C_{1}^{(0)} \* (s^2 - t \* \zu + D/\beta^2 - 2 \* m^2 \* u \* z_{2u}/(s
\beta^2))
                - C_{3u}^{(0)} \* 2 \* u \* (m^2 + z_{2u})
\nonumber \\   
& &\mbox{\hspn}
                + C_{4}^{(0)} \* (D + s^2 - t \* u)
                - D_{2u}^{(0)} \* u \* (D + s^2 - t \* u) \Big]/D
\nonumber
\; ,
\quad
\\[2mm]
\label{eq:QB4}
  Q_{17} &\! =\! &
            16 \* m^2/(s \* \beta^2)
               + B_{1u}^{(0)} \* 2
               - B_{5}^{(0)} \* 2/\beta^2
               - C_{1}^{(0)} \* (4 \* u \* (D + m^2 \* u) + s^2 \* U/\beta^2 + 4 \* m^2 \*
u^2/\beta^2)/D
               + C_{3u}^{(0)} \* 2 \* u \* U \* z_{2u}/D
\nonumber \\
& &\mbox{\hspn}
               - C_{4}^{(0)} \* z_{2u} \* (D + u^2)/D
               + D_{2u}^{(0)} \* u \* z_{2u} \* (D + u^2)/D
\nonumber
\; ,
\quad
\\[2mm]
\label{eq:QB5} 
  Q_{18} &\! =\! &
          \Big[ 8 \* m^2 \* (4 \* s/u - 1 - 5 \* \zu/(s \* \beta^2))
               - B_{1u}^{(0)} \* 2 \* (5 \* D - 3 \* m^2 \* t + t \* u)/u
               - B_{5}^{(0)} \* 2 \* (4 \* t - 5 \* m^2 \* z_{2u}/(s \* \beta^2))
\nonumber \\
& &\mbox{\hspn}
               + C_{1}^{(0)} \* (4 \* s^2 + 2 \* (4 \* s + 5 \* u) \* U + 5 \* m^2 \*
z_{2u}/\beta^2)
               - C_{3u}^{(0)} \* 2 \* u \* (5 \* m^2 + 4 \* z_{2u})
               + C_{4}^{(0)} \* (5 \* m^2 \* s + 4 \* s \* z_{2u} + 10 \* u^2)
\nonumber \\
& &\mbox{\hspn}
               - D_{2u}^{(0)} \* u \* (5 \* m^2 \* s + 4 \* s \* z_{2u} + 10 \* u^2) \Big]/D
\nonumber
\; .
\end{eqnarray}
\end{widetext}


\end{document}